\documentclass[pra,twocolumn,superscriptaddress,showpacs,floatfix]{revtex4}
\usepackage{amsmath}
\usepackage{paralist}
\usepackage{amssymb}
\usepackage{amsthm}
\usepackage{graphicx}
\usepackage{marvosym}
\usepackage{fancyvrb}
\usepackage{calc}
\usepackage{bbold}
\usepackage{ifthen}

\newcommand{\bEq}{\begin{equation}}
\newcommand{\eEq}{\end{equation}}
\newcommand{\ket}[1]{{\left| #1 \right>}}

\newcommand{\on}[1]{\operatorname{#1}}

\renewcommand{\phi}{\varphi}
\renewcommand{\epsilon}{\varepsilon}
\renewcommand{\theta}{\vartheta}

\newtheorem{theorem}{Theorem}
\newtheorem{definition}{Definition}
\newtheorem{corollary}{Corollary}

\newlength{\codelineindent}
\newlength{\codelinelength}
\newcounter{codelinectr}
\newcommand{\codeline}[2]{%
\setlength\codelineindent{1ex+.7cm*#1} %
\setlength\codelinelength{0.5\textwidth-\codelineindent-1cm} %
\refstepcounter{codelinectr}
{\scriptsize \texttt{\ifthenelse{\thecodelinectr<10}{ }{}\thecodelinectr}} \hspace{\codelineindent}
\parbox[t]{\codelinelength}{%
\raggedright #2%
}\\%
}

\begin{document}

\title{Fast simulation of stabilizer circuits using a graph state representation}

\author{Simon Anders}
\email{sanders@fs.tum.de}
\affiliation{Institut f\"ur Theoretische Physik, Universit\"at Innsbruck, Innsbruck, Austria}
\author{Hans J.\ Briegel}
\affiliation{Institut f\"ur Theoretische Physik, Universit\"at Innsbruck, Innsbruck, Austria}
\affiliation{Institut f\"ur Quantenoptik und Quanteninformation der \"Osterreichischen Akademie der Wissenschaften, Innsbruck, Austria}

\date{December, 2005 (v2)}

\begin{abstract}
According to the Gottesman-Knill theorem, a class of quantum circuits, namely the so-called stabilizer circuits, can be simulated efficiently on a classical computer. We introduce a new algorithm for this task, which is based on the graph-state formalism. It shows significant improvement in comparison to an existing algorithm, given by Gottesman and Aaronson, in terms of speed and of the number of qubits the simulator can handle. We also present an implementation.
\end{abstract}


\maketitle

\section{Introduction}

Protocols in quantum information science often use entangled states of a large number of qubits. A major challenge in the development of such protocols is to actually test them using a classical computer. This is because a straight-forward simulation is typically exponentially slow and hence intractable. Fortunately, the Gottesman-Knill theorem (\cite{Got98}, \cite{NC00}) states that an important subclass of quantum circuits can be simulated efficiently, namely so-called \textit{stabilizer circuits}. These are circuits that use only gates from a restricted subset, the so-called Clifford group. Many techniques in quantum information use only Clifford gates, most importantly the standard algorithms for entanglement purification \cite{BBP+96,DAJ+96,MPP+98,MaSm00,DAB03} and for quantum error correction \cite{Sho95,Ste96,CS96,Ste96b}. Hence, if one wishes to study such networks, one can simulate them numerically.

The usual proof of the Gottesman-Knill theorem (as stated e.~g. in \cite{NC00}) contains an algorithm that can carry out this task in time $\mathcal{O}(N^3)$, where $N$ is the number of qubits. Especially for the applications just mentioned, one is interested in a large $N$: For entanglement purification one might want to study large ensembles of states, and for quantum error correction concatenations of codes. The cubic scaling renders this extremely time-consuming, and a more efficient algorithm should be of great use.

Recently, Aaronson and Gottesman presented such an algorithm (and an implementation of it) in Ref.\ \cite{AaGo04}, whose time and space requirements scale only quadratically with the number of qubits. In the present paper, we further improve on this by presenting an algorithm that for typical applications only requires time and space of $\mathcal{O}(N\log N)$. While Aaronson and Gottesman's simulator, when used on an ordinary desktop computer, can simulate already systems of several thousands of qubits in a reasonable time, we have used our simulator for over a million of qubits. This provides a valuable tool for investigating complex protocols such as our study of multi-party entanglement purification protocols in Ref.\ \cite{KADB05}.

The crucial new ingredient is the use of so-called graph states. Graph states have been introduced in \cite{BrRa00} for the study of entanglement properties of certain multi-qubit systems; they were used as starting point for the one-way quantum computer (i.~e., measurement-base quantum computing) \cite{RBB03}, and found to be suited to give a graphical description of CSS codes (for quantum error correction) \cite{ScWe00}. Graph states take their name from the concept of graphs in mathematics: Each qubit corresponds to a vertex of the graph, and the graph's edges indicate which qubits have interacted (see below for details). 

There is an intimate correspondence between stabilizer states (the class of states that can appear in a stabilizer circuit) and graph states: Not only is every graph state a stabilizer state, but also every stabilizer state is equivalent to a graph state in the following sense: Any stabilizer state can be transformed to a graph state by applying a tensor product of local Clifford (LC) operations \cite{Sch01,GKR02,NDM03}. We shall call these local Clifford operators the \textit{vertex operators} (VOPs).

To represent a stabilizer state in computer memory, one stores its tableau of stabilizer operators, which is an $N\times N$ matrix of Pauli operators and hence takes space of order $\mathcal{O}(N^2)$ (see below for details). Gottesman and Aaronson's simulator extends this matrix by another matrix of the same size (which they call the destabilizer tableau), so that their simulator has space complexity $\mathcal{O}(N^2)$. A graph state, on the other hand, is described by a mathematical graph, which, for reasons argued later, only needs space of $\mathcal{O}(N\log N)$ in typical applications. Hence, much larger systems can be represented in memory, if one describes them as graph states, supplemented with the list of VOPs. However, we also need efficient ways to calculate how this representation changes, when the represented state is measured or undergoes a Clifford gate application. The effect of measurements has been extensively studied in \cite{HEB03}, and gate application is what we will study in this paper, so that we can then assemble both to a simulation algorithm.

This paper is organized as follows: We first review the stabilizer formalism, the Gottesman-Knill theorem, and the graph state formalism in Section \ref{states}. There, we will also explain our representation in detail. Section \ref{gates} explains how the state representation changes when Clifford gates are applied. This is the main result and the most technical part of the paper. For the simulation of measurements, we can rely on the studies of Ref. \cite{HEB03}, which are reviewed and applied for our purpose in Section \ref{measurements}. Having exposed all parts of the simulator algorithm, we continue by presenting our implementation of it. A reader who only wishes to use our simulator and is not interested in its internals may want to read only this section. Section \ref{performance} assesses the time requirements of the algorithm's components described in Sections \ref{gates} and \ref{measurements} in order to prove our claim of superior scaling of performance. We finish with a conclusion (Section \ref{conclusion}).

\section{Stabilizer and Graph States} \label{states}

We start by explaining the concepts mentioned in the introduction in a formal manner.

\begin{definition}
The \textit{Clifford group} $\mathcal{C}_N$ on $N$ qubits is defined as the normalizer of the \textit{Pauli group} $\mathcal{P}_N$:
\begin{multline}\mathcal{C}_N = \left\{U\in SU(2^N) \mid UPU^\dagger \in \mathcal{P}_N\quad \forall P\in\mathcal{P}_N\right\},\\
\mathcal{P}_N = \{\pm 1, \pm i\} \cdot \mathcal \{I, X, Y, Z\}^{\otimes N},
\end{multline}
where $I$ is the identity and $X$, $Y$, and $Z$ are the usual Pauli matrices.
\end{definition}

The Clifford group can be generated by three elementary gates (see e.~g. \cite{NC00}): the Hadamard gate $H$, the $\frac{\pi}{4}$ phase rotation $S$, and a two-qubit gate, either the controlled \textsc{not} gate $\Lambda X$, or the controlled phase gate $\Lambda Z$:
\[H = \frac{1}{\sqrt{2}}\left(\begin{array}{cc} 1 & 1 \\ 1 & -1 \end{array}\right) \qquad
S = \left(\begin{array}{cc} 1 & 0 \\ 0 & i \end{array}\right) \]
\bEq\Lambda X = \left(\begin{array}{cccc} 1 & 0 & 0 & 0 \\ 0 & 1 & 0 & 0 \\ 0 & 0 & 0 & 1 \\ 0 & 0 & 1 & 0 \\  \end{array}\right) \qquad
\Lambda Z = \left(\begin{array}{cccc} 1 & 0 & 0 & 0 \\ 0 & 1 & 0 & 0 \\ 0 & 0 & 1 & 0 \\ 0 & 0 & 0 & -1 \\  \end{array}\right) \label{matrices}\eEq

The significance of the Clifford group is due to the Gottesman-Knill theorem (\cite{Got98}, see also \cite{NC00}):

\begin{theorem} A quantum circuit using only the following elements (called a \textit{stabilizer circuit}) can be simulated efficiently on a classical computer:
\begin{compactitem}
\item preparation of qubits in computational basis states
\item quantum gates from the Clifford group
\item measurements in the computational basis
\end{compactitem}
\end{theorem}

The proof of the theorem is simple after one introduces the notion of stabilizer states \cite{Got97}:

\begin{definition}
An $N$-qubit state $\ket{\psi}$ is called a \textit{stabilizer state} if it is the unique eigenstate with eigenvalue +1 of $N$ commuting multi-local Pauli operators $P_a$ (called the stabilizer generators):
\[ P_a\ket{\psi} = \ket{\psi},\quad P_a \in \mathcal{P}_N,\quad a=1,\dots,N\] 
\end{definition}
(These $N$ operators generate an Abelian group, the \textit{stabilizer}, of $2^N$ Pauli operators that all satisfy this stabilization equation.)

Computational basis states are stabilizer states. Furthermore, if a \textit{Clifford} gate $U$ acts on a stabilizer state $\ket{\psi}$, the new state $U\ket{\psi}$ is a stabilizer state with generators $UP_iU^\dagger\in\mathcal{P}_N$. Hence, the state in a stabilizer circuit can always be described by the \textit{stabilizer tableau}, which is a matrix of $N\times N$ operators from $\{I,X,Y,Z\}$ (where each row is preceded by a sign factor). The effect of an $n$-qubit gate can then be determined by updating $nN$ elements of the matrix, which is an efficient procedure.

Instead of on the stabilizer tableau, we shall base our state representation on graph states:

\begin{definition}
An $N$-qubit graph state $\ket{G}$ is a quantum state associated with a mathematical graph $G=(V,E)$, whose $|V|=N$ vertices correspond to the $N$ qubits, while the edges $E$ describe quantum correlations, in the sense that $\ket{G}$ is the unique state satisfying the $N$ eigenvalue equations
\begin{multline} 
K_G^{(a)} \ket{G} = \ket{G},\quad a\in V,\\ 
\text{with } K_G^{(a)} = \sigma_x^{(a)} \prod_{b\in\operatorname{ngbh} a} \sigma_z^{(b)} =: X_a \prod_{b\in\operatorname{ngbh} a} Z_b, \label{corrop}
\end{multline}
where $\on{ngbh} a := \left\{b\mid \{a,b\}\in E\right\}$ is the set of vertices adjacent to $a$ \cite{RBB03, BrRa00, ScWe00}.
\end{definition}

The following theorem states that the edges of the graph can be associated with phase gate interactions between the corresponding qubits:
\begin{theorem} 
If one starts with the state $\ket{+}^{\otimes N} = \prod_{a\in V}H_a \ket{00\dots 0}$ one can easily construct $\ket{G}$ by applying $\Lambda Z$ on all pairs of neighboring qubits:
\bEq \ket{G} = \left(\prod_{\{a,b\}\in E} \Lambda Z_{ab}\right) \left(\prod_{a\in V} H_a\right) \ket{0}^{\otimes N} \label{constr_graph}\eEq
\end{theorem}
(Proof: Insert Eq.~(\ref{constr_graph}) into Eq.~(\ref{corrop}) \cite{HEB03}.)

As the operators $K_G^{(a)}$ belong to the Pauli group, all graph states are stabilizer states, and so are the states which we get by applying \textit{local} Clifford operators $C\in\mathcal{C}_1$ to $\ket{G}$. For such states, we introduce the notation
\bEq \ket{G; \underline{C}} := \ket{G; C_1, C_2, \dots, C_N} := \bigotimes_{i=1}^{N} C_i \ket{G} \label{graphnot}\eEq

It has been shown that all stabilizer states can be brought into this form \cite{Sch01,GKR02,NDM03}, i.~e. any stabilizer state is LC-equivalent to a graph state. (We call two states LC-equivalent if one can be transformed into the other by applying a tensor product of local Clifford operators.) Finding the graph state that is LC-equivalent to a stabilizer state given by a tableau can be done by a sort of Gaussian elimination as explained in \cite{NDM03}.

\begin{figure}
\begin{center}
\begin{minipage}[t]{4cm}
(a)
\begin{tabular}[t]{ccccc}
  & 1 & 2 & 3 & 4 \\ \hline\hline
\it
$+$ & $Z$ & $Z$ & $X$ & $I$ \\
$+$ & $X$ & $X$ & $X$ & $I$ \\
$-$ & $X$ & $Z$ & $Y$ & $Z$ \\
$+$ & $I$ & $I$ & $X$ & $Y$
\end{tabular}
\end{minipage}
\hspace{.5cm}
(b)\raisebox{-2cm}{\parbox[b]{3cm}{\includegraphics[width=2.6cm]{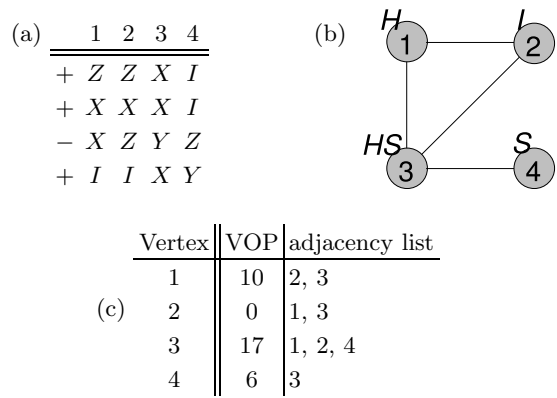}}}\\[3ex]

(c)
\begin{minipage}[t]{4cm}
\begin{tabular}{c||c|l}
Vertex & VOP & adjacency list \\ \hline
1 & 10 & 2, 3 \\
2 &  0 & 1, 3 \\
3 & 17 & 1, 2, 4 \\
4 &  6 & 3 \\
\end{tabular}
\end{minipage}
\caption{A stabilizer state $\ket{\psi}$ represented in different ways: (a) as stabilizer tableau, i.~e. the state is stabilized by the group of Pauli operators generated by the operators in the 4 rows. This representation needs space $\mathcal{O}(N^2)$ for $N$ qubits. (b), (c) as LC-equivalence to a graph state. (b) shows the graph, with the VOPs given by their decomposition into the group generators $\{H,S\}$. (c) is the data structure that represents (b) in our algorithm. The VOPs are now specified using numbers between 0 and 23 (which enumerate the $|\mathcal{C}_1|=24$ LC operators). Here, we need space $\mathcal{O}(N\overline{d})$, where $\overline{d}$ is the average vertex degree, i.~e. the average length of the adjacency lists. Writing $G$ for the graph in (b), we can use the notation of Eq.~(\ref{graphnot}) and write $\ket{\psi} = \ket{G;H,I,HS,S}$.} \label{figrepr}
\end{center}
\end{figure}

This is what we shall use to represent the current quantum state in the memory of our simulator. Fig.~\ref{figrepr} shows for an example state the tableau representation that is usually employed (and also used by CHP, albeit in a modified form) and our representation. The tableau representation requires space of order $\mathcal{O}(N^2)$. We store the graph in adjacency list form (i.~e., for each vertex, a list of its neighbors is stored), which needs space of order $\mathcal{O}(N\overline{d})$, where $\overline{d}$ is the average vertex degree (number of neighbors) in the graph. We also store a list of the $N$ local Clifford operators $C_1,\dots,C_N$, which transform the graph state $\ket{G}$ into the stabilizer state $\ket{G;\underline{C}}$. We call these operators the vertex operators (VOPs). As there are only 24 elements in the local Clifford group, each VOP is represented as a number in $0,\dots,23$. The scheme to enumerate the 24 operators will be described in \cite{And05}. Note that we can disregard global phases of the VOPs as they only lead to a global phase of the full state of the simulator.

As we shall see later, we may typically assume that $\overline{d}=\mathcal{O}(\log N)$. Hence, our representation needs considerably less space in memory than a tableau, namely $\mathcal{O}(N \log N)$, including $\mathcal{O}(N)$ for the VOP list.

The Gaussian elimination needed to transform a stabilizer tableau to its graph state representation is slow (time complexity $\mathcal{O}(N^3)$), and so we should better not use it in our simulator. But usually, one starts with the initial state $\ket{0}^{\otimes N}$, and if we write this state already in graph state form, the tableau representation is never used at all.

From Eq.~(\ref{constr_graph}), it is clear that the initial state can be written as a graph with no edges and Hadamard gates acting on all vertices:
\[\ket{0}^{\otimes N} = \ket{(\{1,\dots,N\}, \{\});H,\dots,H}.\]

\section{Gates}\label{gates}

When the simulator is asked to simulate a Clifford gate, the current stabilizer state is changed and its graph representation has to be updated to correctly reflect the action of the gate. How to do this, is the main technical result of this paper.

\subsubsection{Single-qubit gates} In the graph representation, applying local (single-qubit) Clifford gates becomes trivial: if $C\in\mathcal{C}_1$ is applied to qubit $a$, we replace this qubit's VOP $C_a$ by $CC_a$.\medskip

\subsubsection{Two-qubit gates} It is sufficient if the simulator is  capable to simulate a single multi-qubit gate: As the entire Clifford group is generated, e.~g., by $H$, $S$, and $\Lambda Z$, all gates can be constructed by concatenating these. We chose to implement $\Lambda Z$, the phase gate, as this is (because of its role in Eq.~(\ref{constr_graph})) most natural for the graph-state formalism. 

In the following discussion, the two qubits onto which the phase gate acts, are called the \textit{operand vertices} and denoted with $a$ and $b$. All other qubits are called \textit{non-operand vertices} and denoted $c,d,\dots$.

To solve the task, we have to distinguish several cases.

\smallskip \textbf{Case 1.} \textit{The VOPs of both operand vertices are in $\mathcal{Z}$,} where $\mathcal{Z} := \{I, Z, S, S^\dagger\}$ denotes the set of those four local Clifford operators that commute with $\Lambda Z$ (the other 20 operators do not). In this case, applying the phase gate is simple: We use the fact that (due to Eq.~(\ref{constr_graph})) applying a phase gate on a graph state just toggles an edge:
\[\Lambda Z_{ab} \ket{(V,E)} = \ket{(V,E\vartriangle\{\{a,b\}\})},\label{phasegateongraph}\]
where $\vartriangle$ denotes the symmetric set difference $A\vartriangle B:=(A \cup B) \backslash (A \cap B)$, i.~e. the edge $\{a,b\}$ is added to the graph if is was not present before, otherwise it is removed. 

\smallskip \textbf{Case 2.} \textit{The VOP of at least one of the operand vertices is not in $\mathcal{Z}$.} In this case, just toggling the edge is not allowed because the $\Lambda Z_{ab}$ cannot be moved past the non-$\mathcal{Z}$ VOP. But there is a way to change the VOPs without changing the state, which works in the following case:

\smallskip \textbf{Sub-case 2.2.} \textit{Both operand vertices have non-operand neighbors}. Here, the following operation will help:

\begin{definition} The operation of \textit{local complementation} about a vertex $a$ of a graph $G = (V, E)$, denoted $L_a$, is the operation that inverts the subgraph induced by the neighborhood of $v$:
\[L_a(V,E) = (V, E \vartriangle \{\{b,c\}|b,c\in\operatorname{ngbh} a\})\] \label{defloccompl}
\end{definition}

This operation transforms the state into a local-Clifford equivalent one, as the following theorem, taken from \cite{HEB03, NDM03}, asserts:

\begin{theorem} Applying the local complementation $L_a$ onto a graph $G$ yields a state $\ket{L_a G} = U\ket{G}$, with the multi-local unitary
\[ U = \sqrt{-iX_a} \prod_{b\in\on{ngbh} a} \sqrt{iZ_b} \propto \sqrt{K_G^{(a)}} .\]
\label{lurule} \end{theorem}

Note that the operator $\sqrt{iZ}$ is related to the phase operator $S$ of Eq.~(\ref{matrices}): $\sqrt{iZ} = e^{i\frac{\pi}{4}}S^\dagger$, and $\sqrt{iX}=\sqrt{-iX}^\dagger= \frac{1}{\sqrt{2}}\left(\begin{array}{cc} 1 & -i \\ -i & 1 \end{array}\right)$.

An obvious consequence of Theorem \ref{lurule} is the following.
\begin{corollary} A state $\ket{G;\underline{C}}$ is invariant under application of $L_a$ to $G$, followed by an updating of $C$ according to
\bEq
C_b \mapsto \left\{\begin{array}{ll} 
C_b \sqrt{iX} & \text{for } b = a \\
C_b \sqrt{-iZ} & \text{for } b \in \operatorname{ngbh} a \\
C_b & \text{otherwise}
\end{array}\right. .\label{invbyprod}
\eEq
\end{corollary}

Now note that the local Clifford group is generated not only by $S$ and $H$ but also by $\sqrt{-iX}$ and $\sqrt{iZ}$, the Hermitian adjoints of the operators right-multiplied to the VOPs in Eq.~(\ref{invbyprod}). Our simulator has a look-up table that spells out every local Clifford operator as a product of --as it turns out, at most 5-- of these two operators, times a disregarded global phase. For example, the table's line for $H$ reads:
\bEq H\propto\sqrt{-iX}\sqrt{iZ}\sqrt{iZ}\sqrt{iZ}\sqrt{-iX}.\label{decompex}\eEq

This allows us now to reduce the VOP $C_a$ of any \textit{non-isolated} vertex $a$ to the identity $I$ by proceeding as follows: The decomposition of $C_a$ taken from the look-up table is read from right to left. When a factor $\sqrt{-iX}$ is read we do a local complementation about $a$. This does not change the state if the correction of Eq.~(\ref{invbyprod}) is applied, which right-multiplies a factor $\sqrt{iX}$ to $C_a$. This factor $\sqrt{iX}$ cancels with the factor $\sqrt{-iX}$ at the right-hand end of $C_a$'s decomposition, so that we now have a VOP with a shorter decomposition.

If the right-most operator of the decomposition is $\sqrt{iZ}$ we do a local complementation about an arbitrarily chosen neighbor of $a$, called $a$'s ``swapping partner''. Now, the correction operation will lead to a factor $S$ being right-multiplied to $C_a$, again shortening the decomposition.

Note that a local complementation about $a$ never changes the edges incident on $a$ and hence, if $a$ was non-isolated in the beginning of the procedure, it will stay so. This is important, as only a non-isolated vertex can have a swapping partner. Hence, the procedure can be iterated, and (as the decompositions have a maximum length of 5) after at most 5 iterations, we are left with the identity $I$ as VOP. 

We apply the described ``VOP reduction procedure'' to both operand vertices. After that, both vertices are the identity, and we can proceed as in Case 1.

One might wonder, however, whether the use of the VOP reduction procedure on the second operand vertex $b$ spoils the reduction of the VOP of the first operand $a$. After all, $a$ could be a neighbor of $b$ or of the swapping partner $c$ of $b$. Then, if a local complementation $L_b$ or $L_c$ is performed, the compensation according to Eq.\ (\ref{invbyprod}) changes the neighborhood of $b$ and $c$ (which include $a$). But note that a neighbor of the inversion center only gets a factor $\sqrt{-iZ}\propto S^\dagger$. As $S^\dagger$ generates $\mathcal{Z}$, this means that after the reduction of $b$, the VOP of $a$ might be no longer the identity but it is still an element of $\mathcal{Z}$, and we are allowed to go on with Case 1.

But what happens, if one of the vertices does not have a non-operand neighbor, that could serve as swapping partner? This is the next Sub-case.

\medskip \textbf{Sub-case 2.2.} \textit{At least one of the operand vertices is isolated or only connected to the other operand vertex.} We first assume that the other vertex is non-connected in the same sense:

\medskip \textbf{Sub-sub-case 2.2.1.} \textit{Both operand vertices are either completely isolated, or only connected with each other.} Then, we can ignore all other vertices and have to study only a finite, rather small number of possible states.

Let us denote by $\bullet\,\,\bullet$ the 2-vertex graph with no edges, and by $\bullet\!\!-\!\!\bullet$ the 2-vertex graph with one edge. There are only very few possible 2-qubit stabilizer states, namely those in
\bEq\mathcal{S}_2 := \left\{\ket{G;C_1,C_2}\mid G\in\{\bullet\,\,\bullet, \bullet\!\!-\!\!\bullet\}, C_1, C_2\in\mathcal{C}_1\right\}.\label{S2}\eEq
Of course, many of the assignments in the r.h.s describe the same state, such that $|\mathcal{S}_2| < 2\cdot 24^2$.
Remember that the phase gate $\Lambda Z_{1,2}$ (being a Clifford operator) maps $\mathcal{S}_2$ bijectively onto itself.

The function table of $\Lambda Z_{1,2}|_{\mathcal{S}_2}: \ket{G;C_1,C_2}\mapsto\ket{G';C'_1,C'_2}$ can easily be computed in advance (we did it with Mathematica) and hard-coded into the simulator as a look-up table. This table contains $2\cdot 24^2$ lines such as 
\bEq \ket{\bullet\,\,\bullet, C_{[13]}, C_{[2]}}\mapsto\ket{\bullet\!\!-\!\!\bullet, C_{[0]}, C_{[2]}},\label{phasetableex}\eEq
where the $C_{[i]} (i=0,\dots,23)$ are the Clifford operators in the enumeration detailed in \cite{And05} (e.~g.\ $C_{[0]}=I$, $C_{[2]}=Y$).

Note that many of the assignments to $C_1$ and $C_2$ in Eq.~(\ref{S2}) describe the same state. Hence, we have a choice in the operators $C'_1$, $C'_2$ with which we represent the results of the phase gate in the look-up table. It turns out (by inspection of all the possibilities) that we can always choose the operators such that the following constraint is fulfilled:

\smallskip \textbf{Constraint 1.} \textit{If $C_1 (C_2)\in \mathcal{Z}$, choose $C'_1, C'_2$ such that again $C'_1 (C'_2)\in\mathcal{Z}$.}

The use of this will become clear soon.

\medskip\textbf{Sub-case 2.2.2.} We are left with one last case, namely that one vertex, let it be $a$, is connected with non-operand neighbors, but the other vertex $b$ is not, i.~e. has either no neighbors or only $a$ as neighbor. Then, we proceed as follows: We use iterated local complementations to reduce $C_a$ to $I$. After that, we may use the look-up table as in Sub-sub-case 2.2.1. That this is allowed even though $a$ is connected to a non-operand vertex is shown in the following: First note that the state after the reduction of $C_a$ to $I$ can be written (following Eq.~(\ref{graphnot})) as
\begin{multline}
\ket{(V,E);\underline{C}} = \prod_{c\in V} C_c \prod_{\{c,d\}\in E} \Lambda Z_{cd}\, \ket{++\dots+} \nonumber \\
= \underbrace{\prod_{\substack{c\in\\ V\backslash\{a,b\}}} C_c \prod_{\substack{\{c,d\}\in\\ E\backslash\{\{a,b\}\}}} \Lambda Z_{cd}}_{\substack{C_b\text{ and }\Lambda Z_{ab}\\ \text{commute with this}}}\,
\underbrace{\vphantom{\prod_{\substack{A\\A}}}C_b \left(\Lambda Z_{ab}\right)^\zeta \ket{++\dots+}}_{\substack{=\ket{+}^{\otimes{N-2}} \otimes \ket{\phi}_{ab}\\\text{with }\ket{\phi}\in\mathcal{S}_2\\(\star)}}
\end{multline}
(where $\zeta=0,1$ indicates whether $\{a,b\}\in E$). Observe that $C_b$ has been moved past the operators $\Lambda Z_{cd}$. This is allowed because none of the $\Lambda Z_{cd}$ acts on $b$

We now apply $\Lambda Z_{ab}$ to this state. $\Lambda Z_{ab}$ can be moved through all the phase gates and vertex operators above the left brace so that it stands right in front of the $\mathcal{S}_2$ state $\ket{\phi}_{ab}$ which is separated from the rest. Thus, the table (\ref{phasetableex}) from Sub-sub-case 2.2.1 may be used. (This would not be the case if, in the state above the brace marked with ``$(\star)$'', the two operand vertices were still entangled with other qubits.) The table look-up will give new operators $C'_a, C'_b$ and a new $\zeta'$, so that the new state has the following form:

\begin{multline}
\Lambda Z_{ab} \ket{(V,E);\underline{C}} =\\
\prod_{\substack{c\in\\ V\backslash\{a,b\}}} C_c \prod_{\substack{\{c,d\}\in\\ E\backslash\{\{a,b\}\}}} \,\,\, \Lambda Z_{cd}\,\, C'_a C'_b \left(\Lambda Z_{ab}\right)^{\zeta'} \ket{++\dots+} \end{multline}

For this to be a state in our usual $\ket{G; \underline{C}}$ form (\ref{graphnot}), the two operators $C'_a$ and $C'_b$ have to moved to the left, through the $\Lambda Z_{cd}$. For $C'_b$, this is no problem, as $b$ was assumed to be either isolated or connected only to $a$, so that $C'_b$ commutes with $\prod_{\{c,d\}\in E\backslash\{\{a,b\}\}} \Lambda Z_{cd}$, as the latter operator does not act on $b$. The vertex $a$, however, has connections to non-operand neighbors, so that some of the $\Lambda Z_{cd}$ act on it. We may move it only if $C'_a\in\mathcal{Z}$ (as this means that it commutes with $\Lambda Z$). Luckily, due to Constraint 1 imposed above, we can be sure that $C'_a\in\mathcal{Z}$, because $C_a=I\in\mathcal{Z}$.
\medskip

Listing 1 shows in pseudo-code how these results can be used to actually implement the controlled phase gate $\Lambda Z$.

\begin{table}
\begin{flushleft}
\codeline{0}{\texttt{cphase} (vertex $a$, vertex $b$):}
\codeline{1}{\textbf{if} $\operatorname{ngbh} a \backslash \{b\} \neq \{\}$:}
\codeline{2}{\texttt{remove\_VOP} ($a,b$)}
\codeline{1}{\textbf{end if}}
\codeline{1}{\textbf{if} $\operatorname{ngbh} b \backslash \{a\} \neq \{\}$:}
\codeline{2}{\texttt{remove\_VOP} ($b,a$)}
\codeline{1}{\textbf{end if}}
\codeline{1}{\textit{[It may happen that the condition in line 2 has not been fulfilled then, but is now due to the effect of line 5. So we check again:]}}
\codeline{1}{\textbf{if} $\operatorname{ngbh} a \backslash \{b\} \neq \{\}$:}
\codeline{2}{\texttt{remove\_VOP} ($a,b$)}
\codeline{1}{\textbf{end if}}
\codeline{1}{\textit{[Now we can be sure that the the condition $(\operatorname{ngbh} c \backslash \{a,b\} = \{\}$ \textbf{\rm or} $\text{\texttt{VOP}}[c]\in\mathcal{Z})$ is fulfilled for $c=a,b$ and we may use the lookup table (cf.\ Eq.~(\ref{phasetableex})).]}}
\codeline{1}{\textbf{if} $\{a,b\}\in E:$}
\codeline{2}{\texttt{edge} $\leftarrow$ \textbf{true}}
\codeline{1}{\textbf{else:}}
\codeline{2}{\texttt{edge} $\leftarrow$ \textbf{false}}
\codeline{1}{\textbf{end if}}
\codeline{1}{$(\text{\texttt{edge}}, \text{\texttt{VOP}}[a], \text{\texttt{VOP}}[b]) \leftarrow\quad~\text{\texttt{cphase\_table}}[\text{\texttt{edge}}, \text{\texttt{VOP}}[a], \text{\texttt{VOP}}[b]]$}
\codeline{0}{}
\codeline{0}{\texttt{remove\_VOP} (vertex $a$, vertex $b$):}
\codeline{0}{\textit{[This reduces \texttt{VOP}$[a]$ to $I$, avoiding (if possible) to use $b$ as swapping partner.]}}
\codeline{1}{\textit{[First, we choose a swapping partner $c$.]}}
\codeline{1}{\textbf{if} $\operatorname{ngbh} a\backslash\{b\} \neq\{\}$:}
\codeline{2}{$c\leftarrow \text{ any element of } \operatorname{ngbh} a\backslash\{b\}$}
\codeline{1}{\textbf{else:}}
\codeline{2}{$c\leftarrow b$}
\codeline{1}{\textbf{end if}}
\codeline{1}{$d\leftarrow$\texttt{ decomposition\_lookup\_table} $[a]$}
\codeline{1}{\textit{[$c$ contains now a decomposition such as Eq. (\ref{decompex})]}}
\codeline{1}{\textbf{for} $v$ \textbf{from} last factor of $d$ \textbf{to} first factor of $d$}
\codeline{2}{\textbf{if} $v=\sqrt{-iX}$:}
\codeline{3}{\texttt{local\_complementation} ($a$)}
\codeline{2}{\textbf{else:} \textit (this means that $v=\sqrt{iZ}$)}
\codeline{3}{\texttt{local\_complementation} ($b$)}
\codeline{2}{\textbf{end if}}
\codeline{1}{\textit{[Now, \texttt{VOP}$[a] = I$}.]}
\codeline{0}{}
\codeline{0}{\texttt{local\_complementation} (vertex $a$)}
\codeline{0}{\textit{[performs the operation specified in Definition \ref{defloccompl}]}}
\codeline{1}{$n_v \leftarrow \on{ngbh} v$}
\codeline{1}{\textbf{for} $i\in n_v$:}
\codeline{2}{\textbf{for} $j\in n_v$:}
\codeline{3}{\textbf{if} $i<j$:}
\codeline{4}{\textbf{if} $(i,j)\in E$:}
\codeline{5}{remove edge $(i,j)$}
\codeline{4}{\textbf{else}:}
\codeline{5}{add edge $(i,j)$}
\codeline{4}{\textbf{end if}}
\codeline{3}{\textbf{end if}}
\codeline{2}{\textbf{end for}}
\codeline{2}{$\text{\texttt{VOP}}[i]\leftarrow\text{\texttt{VOP}}[i]\sqrt{-iZ}$}
\codeline{1}{$\text{\texttt{VOP}}[v]\leftarrow\text{\texttt{VOP}}[v]\sqrt{iX}$}
\codeline{1}{\textbf{end for}}
\end{flushleft}

LISTING 1: Pseudo-code for controlled phase gate ($\Lambda Z$) acting on vertices $a$ and $b$ (\texttt{cphase}), and for the two auxiliary routines \texttt{remove\_VOP} and \texttt{local\_complementation}.
\end{table}

\section{Measurements} \label{measurements}

In a stabilizer circuit, the simulator may be asked at any point to simulate the measurement of a qubit in the computational basis. How the outcome of the measurement is determined, and how the graph representation has to be updated in order to then represent the post-measurement state will be explained in the following.

To measure a qubit $a$ of a state $\ket{G, \underline{C}}$ in the computational basis means to measure the qubit in the underlying graph state $\ket{G}$ in one of the 3 Pauli bases. Writing the measurement outcome as $\zeta$, this means:
\begin{multline}
\frac{I+(-1)^\zeta Z_a}{2}\ket{G, \underline{C}} =  \left(\prod_{b\in V\backslash\{a\}} C_b\right) \frac{I +(-1)^\zeta Z_a}{2} C_a \ket{G} \\= \left(\prod_{b\in V\backslash\{a\}} C_b\right)  C_a \frac{I+(-1)^\zeta C_a^\dagger Z_aC_a}{2} \ket{G}
\end{multline}
As $C_a$ is a Clifford operator, $P_a:=C_a^\dagger Z_aC_a\in\{X_a,Y_a,Z_a, -X_a,-Y_a,-Z_a\}$. Thus, in order to measure qubit $a$ of $\ket{G,\underline{C}}$ in the \textit{computational} basis, we measure the observable $P_a$ on $\ket{G}$. Note that in case that $P_a$ is the negative of a Pauli operator, the measurement result $\zeta$ to be reported by the simulator is the complement of $\vphantom{\smash[b]{\Big(}}\tilde\zeta$, the result given by the $X$, $Y$ or $Z$ measurement on the underlying graph state $\ket{G}$.

How is the graph $G$ changed and how do the vertex operators have to be modified if the measurement $\frac{I\pm P_a}{2} \ket{G}$ is carried out? This has been worked out in detail in Ref.~\cite{HEB03}, which we now briefly review for the present purpose.

The simplest case is that of $P=\pm Z$. Here, the state changes as follows:
\begin{multline}
\frac{I+(-1)^{\tilde\zeta} Z_a}{2} \ket{(V,E)} = \\
\underbrace{\left(X_a \prod_{b\in\on{ngbh}a}Z_b\right)^{\tilde\zeta} H_a }_{(\star)}\,\ket{(V,E\backslash\{\{a,b\}|b\in\on{ngbh}a\})}.
\end{multline}
The value of $\tilde\zeta$ is chosen at random (using a pseudo-random number generator). To update the simulator state, the VOPs are right-multiplied with the under-braced operators $(\star)$ and the edges incident on $a$ are deleted as indicated in the ket.

A measurement of the $Y$ observable ($P=\pm Y$) requires a complementation of the edges set according to
\[E \mapsto E \vartriangle \left\{ \{b,c\} \mid b,c\in \on{ngbh} a \right\} \]
and a change in the VOPs as follows:
\[C_b \mapsto C_b \sqrt{-iZ}^{(\dagger)} \text{ for } b\in\on{ngbh} a \cup \{a\},\]
where the dagger in parentheses is to be read only for measurement result $\tilde\zeta=1$.

The most complicated case is the $X$ measurement which requires an update of edges and VOPs as follows:
\begin{align}
E \mapsto  E 
&\vartriangle \left\{ \{c,d\} \mid c\in\on{ngbh} b,\, d\in\on{ngbh} a \right\} \nonumber\\
&\vartriangle \left\{ \{c,d\} \mid c,d\in\on{ngbh} b\cap\on{ngbh} a \right\} \nonumber\\
&\vartriangle \left\{ \{b,d\} \mid d\in\on{ngbh} a\backslash\{b\} \right\} \nonumber
\end{align}
\begin{align}
C_c\mapsto & \left\{\begin{array}{ll}
C_c Z^{\tilde\zeta} & \text{for } c=a\\[1.2ex]
C_c \sqrt{iY}^{(\dagger)} & \text{for } c=b \quad\text{(read ``${}^\dagger$'' only for $\tilde\zeta=1$)}\\[1.2ex]
C_c Z & \text{for } c \in \left\{ \begin{array}{l} 
\on{ngbh} a \,\backslash\, \on{ngbh} b \,\backslash\, \{b\}\\
\qquad \text{(for $\tilde\zeta=0$)} \\[1.2ex]
\on{ngbh} b \,\backslash\, \on{ngbh} a \,\backslash\, \{a\}\\
\qquad\text{(for $\tilde\zeta=1$)} \end{array}\right. \\[1.2ex]
C_c & \text{otherwise}
\end{array}\right.
\end{align}
Here, $b$ is a vertex chosen arbitrarily from $\on{ngbh} a$ and $\sqrt{iY}=\frac{1}{\sqrt{2}}\left(\begin{array}{cc}1&-1\\1&1\end{array}\right)$.

In all these cases the measurement result is chosen at random. Only in case of the measurement of $P_a=\pm X$ an isolated vertex, the result is always $\tilde\zeta=0$ (which means an actual result of $\zeta=0$ for $P_a=X$ and $\zeta=1$ for $P_a=-X$.)

\section{Implementation} \label{implementation}

The algorithm described above has been implemented in C++ in object-oriented programming style. We have used the GNU Compiler Collection (GCC) \cite{GCCHL} under Linux, but it should be easy to compile the program on other platforms as well \footnote{We use only ISO Standard C++ with one exception: The \texttt{hash\_set} template is used, which is, though not part of the standard, supplied by most modern compilers.}. The implementation is done as a library to allow for easy integration into other projects. We also offer bindings to Python \cite{PytHL}, so that the library can be used by Python programs as well. (This was achieved using SWIG \cite{SWIG}.)

The simulator, called ``GraphSim'' can be downloaded from \cite{AndHL}. 

A detailed documentation of the library is supplied with it. To demonstrate the usage here at least briefly, we give Listing 2 as a simple toy example. It is written in Python, and a complete program.

\begin{table}
\begin{Verbatim}[numbers=left, xleftmargin=2em]
import random
import graphsim

gr = graphsim.GraphRegister (8)

gr.hadamard (4)
gr.hadamard (5)
gr.hadamard (6)
gr.cnot (6, 3)
gr.cnot (6, 1)
gr.cnot (6, 0)
gr.cnot (5, 3)
gr.cnot (5, 2)
gr.cnot (5, 0)
gr.cnot (4, 3)
gr.cnot (4, 2)
gr.cnot (4, 1)

for i in xrange (7):
   gr.cnot (i, 7)

print gr.measure (7)

gr.print_stabilizer ()
\end{Verbatim}
LISTING 2: A simple example in Python
\end{table}

In the example, we start by loading the GraphSim library (Line 2) and then initialize a register of 8 qubits (line 4), which are then all in $\ket{0}$ state. We get an object called ``\texttt{gr}'' of class \texttt{GraphRegister}, which represents the register of qubits. For all following operations, we use the methods of \texttt{gr} to access its functionality. In our example, we simply build up an encoded ``0'' state in the well-known 7-qubit Steane code, which we then measure.

First, we apply Hadamard and \textsc{cnot} gates onto the qubits with number 0 through 6 in order to build up the Steane-encoded ``0'' (Lines 6--17). To check that we did so, we measure the encoded qubit, which is done by using \textsc{cnot} gates to sum up their parity in the eighth qubit (``qubit 7'') (Lines 19, 20). Measuring qubit 7 then gives ``0'', as it should (Line 22).\medskip

For further details on using of the GraphSim library from a C++ or Python program, please see the documentation supplied with the source code \cite{AndHL}.
\medskip

With approximately 1400 lines, GraphSim is complex enough that one cannot take for granted that it faithfully implements the described algorithm without bugs, and testing is necessary. Fortunately, this can be done very conviniently by comparing with Aaronson and Gottesman's ``CHP'' simulator. As these two programs use quite different algorithms to do the same task, it is very unlikely that any bugs, which they might have, produce the \textit{same} false results. Hence, if both programs give the same result, they can reasonably be considered both to be correct.

We set up a script to do random gates and measurements on a set of qubits for millions of iterations. All operations were performed simultaneously with CHP and GraphSim. For measurements whose outcome was chosen at random by CHP, a facility of GraphSim was used that overrides the random choice of measurement outcomes and instead uses a supplied value. For measurements with determined outcome, however, it was checked whether both programs output the same result. Also, every 1000 steps, the stabilizer tableau of GraphSim's state was calculated from its graph representation and compared to CHP's tableau. \footnote{This was done with a Mathematica subroutine which tries to find a row adding and swapping arrangement to transform one tableau into the other.}

After simulation $4\cdot 10^6$ operations on 200 qubits in 18 hours and $2\cdot 10^8$ operations on 20 qubits in 19.7 hours without seeing discrepancies, we are confident that we have exhausted all special cases, so that the two programs can be assumed to always give the same output. As they are based on very different algorithm, this reasonably allows to conclude that they both operate correctly.
\bigskip

\section{Performance} \label{performance}
We now show that our simulator yields the promised performance, i.~e. performs a simulation of $M$ steps in time of order $\mathcal{O}(NdM)$, where $N$ is the number of qubits and $d$ the maximum vertex degree that is encountered during the calculation. Let us go through the different possible simulation steps in order to assess their respective time requirements.

Single-qubit gates are fastest: they only need one look-up in the multiplication table of the local Clifford group (which is hard-coded into the simulator), and are hence of time complexity $\Theta(1)$.

Measurements have a complexity depending on the basis in which they have to be carried out. For a $Z$ measurement, we have to remove the $\on{deg} a$ edges of the measured vertex $a$. As $d$ is the maximum vertex degree that is to be expected within the studied problem, the complexity of a $Z$ measurement is $\mathcal{O}(d)\le\mathcal{O}(N)$ (as $d\le N$).

For a $Y$ and $X$ measurement, we have to do local complementation, which requires dealing with up to $\frac{d(d-1)}{2}$ edges, and hence, the overall complexity of measurements is $\mathcal{O}(d^2)$. 

For the phase gate, the same holds. Here, we need a fixed number (up to 5) of local complementations. Thus, measurements and two-qubit gates take $\mathcal{O}(d^2)$ time.

This would be no improvement to Aaronson and Gottesman's algorithm, if we had $d=\mathcal{O}(N)$. The latter is indeed the case if one applies randomly chosen operations as we did to demonstrate GraphSim's correctness. There, we indeed did not observe any superiority in run-time of GraphSim.

In practice, however, this is quite different. For example, when simulating quantum error correction, one can reasonable assume $d=\mathcal{O}(\log N)$. This is because all QEC schemes avoid to do to many operations on one and the same qubit in a row, as this would spread errors. So, vertex degrees remain small. The same reasoning applies to entanglement purification schemes and, more generally, to all circuits which are designed to be robust against noise.

\begin{figure}
\includegraphics[width=.4\textwidth]{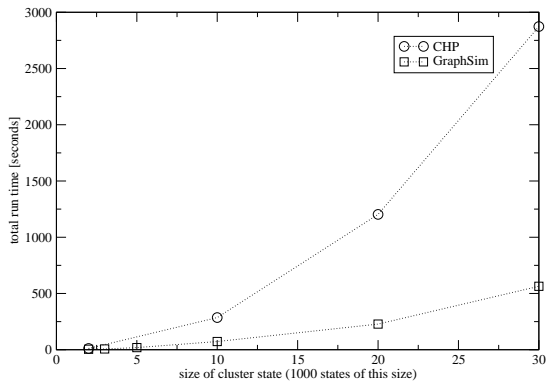}
\caption{Comparison of the performance of CHP and GraphSim. A simulation of entanglement purification was used as sample application. The register has 1000 times the size of the states to hold an ensemble of 1000 states.}
\label{perf}
\end{figure}

\begin{figure}[tb]
\includegraphics[width=.4\textwidth]{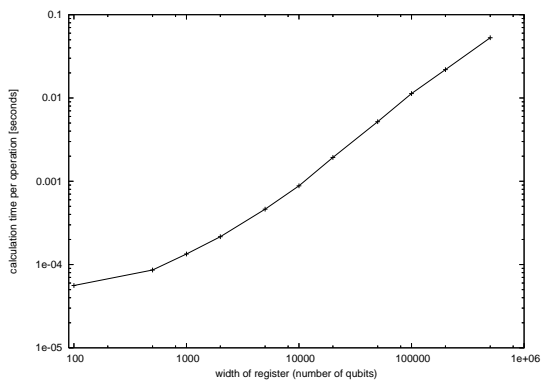}
\caption{Benchmark of GraphSim for very large registers. Entanglement purification --specifically: the purification of 10-qubit cluster states with the protocol of Ref.\ \cite{DAB03}-- was used as sample problem. The register was filled up with cluster states to make a large ensemble, and two protocol steps were simulated. The average time per operation was obtained from the total run-time \cite{footnote}.}
\label{perfb}
\end{figure}

The space complexity is dominated by the space needed to store the quantum state representation. As argued in Section \ref{states}, this requires only space of $\mathcal{O}(N\overline{d})$, where $\overline d$ is the average vertex degree. As explained above, we may expect $\overline d$ (as $d$) to scale sub-linearly with $N$ in typical application, in many applications as $\mathcal{O}(N\log N)$. This is what allows us to handly substantially more qubits than it is possible with the $\mathcal{O}(N^2)$ tableau representation.
\medskip

As a first practical test, we used GraphSim to simulate entanglement purification of cluster states with the protocol of Ref.~\cite{DAB03}. This has been a starting point of a detailed analysis of the communication costs of establishing multipartite entanglement states via noisy channels \cite{KADB05}. Fig.~\ref{perf} demonstrates that GraphSim is indeed suitable for this purpose. Note, that for the right-most data points, the register holds 30,000 qubits. 

As we did a Monte Carlo simulation, we had to loop the calculation very often and still got an output within a few hours. For simulations involving several millions of qubits and a large number of runs, we waited about a week for the results when using eight processors in parallel. We redid some of these calculations in a more controlled testing environment as a benchmark for GraphSim. Fig.~\ref{perfb} shows the results in a log-log plot.

\section{Conclusion} \label{conclusion}

To summarize, we have used recent results on graph states to find a very space-efficient representation of stabilizer states, and determined, how this representation changes under the action of Clifford gates. This can be used to simulate stabilizer circuits more efficiently than previously possible. The gain is not only in simulation speed, but also in the number of manageable qubits. In the latter, at least two orders of magnitude are gained. We have presented an implementation of our simulation algorithm and will soon publish results about entanglement purification which makes use of our new technique.
\vspace{0em}

\begin{acknowledgments}
We would like to thank Marc Hein for most helpful discussions. 

This work was supported in part by the Austrian Science Foundation (FWF),
the Deutsche Forschungsgemeinschaft (DFG), and the European Union
(IST-2001-38877, -39227, OLAQUI, SCALA).
\end{acknowledgments}


\begin{thebibliography}{2}

\bibitem{Got98}{D.~Gottesman, quant-ph/9807006}

\bibitem{NC00}{M.~A.~Nielsen, I.~L.~Chuang: \textit{Quantum Computation and Quantum Information}, Cambridge University Press, 2000}

\bibitem{BBP+96} C.~H.\ Bennett, G.\ Brassard, S.\ Popescu, B.\ Schumacher, J.~A.\ Smolin, W.~K.\ Wootters, Phys.\ Rev.\ Lett. \textbf{76}, 722 (1996).

\bibitem{DAJ+96} D.\ Deutsch, A.\ Ekert, R.\ Jozsa, C.\ Macchiavello, S.\ Popescu, A.\ Sanpera, Phys.\ Rev.\ Lett.\ \textbf{77}, 2818 (1996).

\bibitem{MPP+98} M.\ Murao, M.~B.\ Plenio, S.\ Popescu, V.\ Vedral, and P.~L.\ Knight, Phys.\ Rev.\ A \textbf{57}, R4075 (1998).

\bibitem{MaSm00} E.~N.\ Maneva and J.~A.\ Smolin, In \textit{Quantum Computation
and Quantum Information}, edited by J.\ S.\ J.\ Lomonaco, AMS, Providence, 2002; also quant-ph/0003099.

\bibitem{DAB03}{W.~D\"ur, H.~Aschauer, H.~J.~Briegel, Phys.\ Rev.\ Lett.\ \textbf{91}, 107903 (2003)}

\bibitem{Sho95}{P.~W.~Shor, Phys.~Rev.~A \textbf{52}, 2493 (1995)}

\bibitem{Ste96}{A.~M.~Steane, Phys.~Rev.~Lett.\ \textbf{77}, 793 (1996)}

\bibitem{CS96}{A.~R.\ Calderbank, P.~W.\ Shor, Phys.~Rev.~A \textbf{54}, 1098 (1996)}

\bibitem{Ste96b}{A.~M.~Steane, Proc.~Roy.\ Soc.\ London A \textbf{452}, 2551 (1996)}

\bibitem{AaGo04}{S.~Aaronson, D.~Gottesman, 
Phys.\ Rev.\ A  \textbf{70}, 052328 (2004)}

\bibitem{KADB05}{C.~Kruszynska, S.~Anders, W.~D\"ur, H.~J.~Briegel, quant-ph/0512218}

\bibitem{BrRa00}{H.~J.~Briegel, R.~Rau\ss{}endorf, Phys.\ Rev.\ Lett.\ \textbf{86}, 910 (2001)}

\bibitem{RBB03}{R.~Rau\ss{}endorf, D.~E.~Browne, H.~J.~Briegel, Phys.\ Rev.~A \textbf{68}, 022312 (2003)}

\bibitem{ScWe00}{D.~Schlingemann, R.~F.~Werner, Phys.\ Rev.~A \textbf{65}, 012308 (2002)}

\bibitem{NDM03}{M.~Van~den~Nest, J.~Dehaene, B.~De~Moor, Phys.\ Rev.~A \textbf{69}, 022316 (2004)}

\bibitem{GKR02}{M.~Grassl, A.~Klappenecker, M.~R\"otteler, in \textit{Proceedings of the 2002 IEEE International Symposium on Information Theory (ISIT)}, IEEE, p.~45}

\bibitem{Sch01}{D.~Schlingemann, quant-ph/0111080}


\bibitem{HEB03}{M.~Hein, J.~Eisert, H.~J.~Briegel, Phys.\ Rev.~A \textbf{69}, 062311 (2004)}

\bibitem{Got97}{D.~Gottesman: \textit{Stabilizer Codes and Quantum Error Correction}, Ph.~D. Thesis, California Institute of Technology, 1997. quant-ph/9705052}

\bibitem{And05}{S.~Anders, \textit{A Guide to the Local Clifford Group.} In preparation.}


\bibitem{AndHL}{The described software can be found at\\
\texttt{http://homepage.uibk.ac.at/homepage/c705}\raisebox{-1ex}{\Righttorque}\\
\texttt{\hspace*{2em}/c705213/work/graphsim.html}}

\bibitem{GCCHL}{The GCC Team: \textit{The GNU Compiler Collection}, Software at \texttt{http://gcc.gnu.org}}

\bibitem{PytHL}{\textit{Python}: Programming language developped by Guido van Rossum et al., \texttt{http://www.python.org}}

\bibitem{SWIG}{\textit{SWIG (Simplified Wrapper and Interface Generator):} Software developed by David M. Beazley et al., \texttt{http://www.swig.org}}

\bibitem{footnote}{Giving the time per operation in seconds is of use only when one specifies the machine which has run the code: We used Linux computers with AMD Opteron processors, clocked with 2.2 GHz. Only one the machine's several processors was dedicated to our computation task. The code was compiled using the GNU C++ compiler (version 3.2.3) with 64-bit target and ``O3'' optimization.}

\end{thebibliography}
\end{document}